\documentclass[twocolumn,pra,aps,amsmath,floatfix,showpacs]{revtex4}

\usepackage{graphicx}

\begin{document}
\newcommand {\cs}{$\clubsuit$}
\title{Collective excitations of a trapped boson-fermion mixture
across demixing}

\author{P. Capuzzi, A. Minguzzi, and M.~P. Tosi} 
\affiliation{
NEST-INFM and Classe di Scienze, Scuola Normale Superiore, Piazza dei
Cavalieri 7, I-56126 Pisa, Italy}

\begin{abstract}
We calculate the spectrum of low-lying collective excitations in a
mesoscopic cloud formed by a Bose-Einstein condensate and a
spin-polarized Fermi gas as a function of the boson-fermion
repulsions. The cloud is under isotropic harmonic confinement and its
dynamics is treated in the collisional regime by using the equations
of generalized hydrodynamics with inclusion of surface effects. For
large numbers of bosons we find that, as the cloud moves towards
spatial separation (demixing) with increasing boson-fermion coupling,
the frequencies of a set of collective modes show a softening followed by a
sharp upturn. This behavior permits a clear identification of the
quantum phase transition. We propose a physical interpretation for the
dynamical transition point in a confined mixture, leading to a simple
analytical expression for its location.
\end{abstract}
\maketitle

\newcommand{\rv}{\mathbf{r}}
\newcommand{\drho}{\delta\rho}
\newcommand{\be}{\begin{equation}}
\newcommand{\ee}{\end{equation}}
\newcommand{\bea}{\begin{eqnarray}}
\newcommand{\eea}{\end{eqnarray}}
\newcommand\ignore[1]{{ }}

\section{Introduction}

After the achievement of Bose-Einstein condensation in alkali-atom
gases, advanced techniques are being developed to cool gases of
fermionic isotopes inside magnetic traps. Since the Pauli principle
forbids $s$-wave collisions between spin-polarized fermions, to reach
the degeneracy regime in the fermionic case one must resort to
collisions against a distinguishable species, either bosonic or
fermionic \cite{DeMarco1999a}. Boson-fermion mixtures are currently
being produced and studied in several experiments
\cite{Schreck2001a,Goldwin01a,Hadzibabic02a,Roati02a,Modugno02a,Ferlaino02a}.

  Starting from the work on strongly interacting $^3$He-$^4$He
liquids, cold mixtures have played an important role in the
development and testing of the theory of quantum phase transitions. In
this context trapped mixtures of atomic gases offer a unique
opportunity due to their high diluteness.  Mixtures of a Bose-Einstein
condensate and a degenerate Fermi gas are predicted to have a rich
phase diagram. In the case of attractive boson-fermion interactions,
where the boson-fermion overlap is largest, the boson-induced
fermion-fermion attraction may lead to the formation of a superfluid
state \cite{negative_a}. In the case of repulsive interactions, on the
other hand, the system is expected to undergo spatial separation when
the repulsions overcome the kinetic energy \cite{Molmer_mix}. The
conditions for demixing have been derived for a homogeneous mixture
\cite{Viverit2000a} and also in the experimentally relevant case of a
mixture under harmonic confinement
\cite{Minguzzi00a,Miyakawa00b,Akdeniz02a}.  The static equilibrium
properties of such mixtures across phase separation and the topology
of the particle density profiles in the demixed state have been
studied \cite{Akdeniz02b}. Of course, the transition to the demixed
state in a mesoscopic cloud under confinement is spread out as the
overlap energy between its two components reaches a maximum and then
gradually decreases on further increase of the boson-fermion coupling.

In the mixed state the dynamical properties of a boson-fermion mixture
 have been investigated both for a homogeneous system \cite{Yip2001a}
 and in a mesoscopic cloud under external harmonic confinement
 \cite{dyn_mix}.  The purpose of the present paper is to follow the
 dynamics of a harmonically confined cloud with increasing
 boson-fermion repulsion and to look for a signature of the transition
 to spatial separation in the spectrum of collective modes. We focus
 here on the study of the dynamics in the collisional regime, which at
 high temperature has already been reached in some experiments
 \cite{Ferlaino02a} and can be attained at low temperature in the
 presence of impurities \cite{Amoruso00a}. We solve the hydrodynamic
 equations beyond the Thomas-Fermi approximation, including surface
 effects which are crucial for a proper description of particle
 density fluctuations as the cloud approaches demixing.

The paper is organized as follows. In Sec.\ \ref{sec:theory} we
introduce the specific system that we study and the hydrodynamic
equations which we use throughout the paper. Section \ref{sec:equil}
discusses the equilibrium density profiles that are needed to
evaluate the collective modes of the mixture in Sec.\
\ref{sec:colle}. Finally, Sec.\ \ref{sec:summary} presents a summary
of our results and an outlook towards future developments.

\section{Theoretical Model}
\label{sec:theory}
We consider a dilute fluid composed by two species of alkali atoms,
one fermionic and the other bosonic in a Bose-Einstein condensed
state, 
confined inside a spherical trap at zero temperature. The interactions
between the bosons and between bosons and fermions are described by
contact potentials and are parametrized by the coupling constants
$g_{BB}=4\pi\hbar^2a_{BB}/m_B$ and $g_{BF}=2\pi\hbar^2a_{BF}/m_{r}$ in
terms of the $s$-wave scattering lengths $a_{BB}$ and $a_{BF}$ and of
the masses $m_B$ and $m_F$ of each species, with
$m_{r}=(1/m_B+1/m_F)^{-1}$ being the reduced mass. The fermions are
spin-polarized and are taken as noninteracting, since collisions in
the $s$-wave channel are forbidden by the Pauli principle. In the
following we have chosen $g_{BB}>0$ and $g_{BF}>0$, as for the
$^6$Li-$^7$Li mixture studied in the experiments of Schreck {\it et
al.}~\cite{Schreck2001a}. 

 We describe the dynamics of the system by starting from the equations
of generalized hydrodynamics \cite{March73a} for the particle
densities $\rho_{\sigma}(\rv,t)$ and the current densities
$\mathbf{j}_{\sigma}(\rv,t)$, with $\sigma=B, F$. These equations read
\begin{equation} 
\frac{\partial \rho_{\sigma}}{\partial t} + \nabla \cdot
\mathbf{j}_{\sigma} = 0
\label{Ec:continuidad}
\end{equation}
and 
\begin{equation}
m_{\sigma}\,\frac{\partial \mathbf{j}_{\sigma}}{\partial t} +
\nabla\cdot\Pi^{\sigma} + \rho_{\sigma}\,\nabla\tilde{V}_{\sigma} = 0 . 
\label{Ec:momento}
\end{equation}
Here, $\tilde{V}_{\sigma}$ are the effective mean-field potentials and
$\Pi^{\sigma}$ are the kinetic stress tensors.  In the dilute regime
we may adopt the Hartree-Fock approximation for the effective
potentials,
\begin{equation} 
\tilde{V}_F = V_F^{ext} + g_{BF}\,\rho_B
\label{Ec:potesF}
\end{equation}
and 
\begin{equation}
\tilde{V}_B = V_B^{ext} + g_{BB}\,\rho_B + g_{BF}\,\rho_F,
\label{Ec:potesB}
\end{equation}
where $V_{\sigma}^{ext} = m_{\sigma}\,\omega_{\sigma}^2\,r^2/2$ are
the (isotropic) external trapping potentials. 

The above equations can be closed in the collisional regime, where we
assume a local dependence of the stress tensors on the particle
densities. In the dilute limit the Thomas-Fermi approximation yields
the fermionic stress tensor as the local-density form of the tensor for
the ideal Fermi gas. However, we have added to this Thomas-Fermi form
a surface contribution in the form derived by von Weizs\"acker
\cite{vonWeizsacker35a}, in order to avoid spurious divergences in the
density fluctuations at the classical radius of the cloud. Thus, the
form for $\Pi_{ij}^F$ reads
\begin{eqnarray} 
\Pi^{F}_{ij} &=& \frac{2}{5}\,A\,\rho_F^{5/3}\,\delta_{ij} 
-\frac{\hbar^2}{6\, m_F}\Bigl[\sqrt{\rho_F}\,
\nabla_i\nabla_j\sqrt{\rho_F} \nonumber \\ 
&& - \nabla_i\sqrt{\rho_F}\,
\nabla_j\sqrt{\rho_F}\Bigr]
\label{Ec:PIf}
\end{eqnarray}
where $A=\hbar^2(6\pi^2)^{2/3}/2m_F$. This choice is in agreement with
the general structure of the fermionic stress tensor under harmonic
confinement as demonstrated in Ref.~\cite{Minguzzi2001a}.
In the same approximation the bosonic stress tensor has only the
surface contribution 
\begin{equation} 
\Pi^{B}_{ij} = -\frac{\hbar^2}{2\, m_B}\Bigl[\sqrt{\rho_B}\,
\nabla_i\nabla_j\sqrt{\rho_B} - \nabla_i\sqrt{\rho_B}\,
\nabla_j\sqrt{\rho_B}\Bigr],
\label{Ec:PIb}
\end{equation}
as can also be obtained from the Gross-Pitaevskii equation
\cite{Minguzzi97b}. 
In Eqs.~(\ref{Ec:PIf}) and (\ref{Ec:PIb}) we have left out 
velocity-dependent terms which do not enter linear dynamics.

Using Eqs.~(\ref{Ec:potesF})-(\ref{Ec:PIb}) for the effective
potentials and the kinetic stress tensors we can rewrite the equations
for the current densities as
\begin{equation}
m_{\sigma}\,\frac{\partial\mathbf{j}_{\sigma}}{\partial t} =
\rho_{\sigma}\bigl(\mathbf{F}_{\sigma} - \nabla V_{\sigma} -
g_{BF}\,\nabla \rho_{\bar{\sigma}}\bigr),
\label{Ec:finalsolution}
\end{equation}
where $\bar{\sigma}$ denotes the component different from $\sigma$ and
for convenience we have introduced the forces
\begin{equation} 
\mathbf{F}_B = -
\nabla\left[g_{BB}\,\rho_B - \frac{\hbar^2}{2\,m_B}\,
\frac{\nabla^2\sqrt{\rho_B}}{\sqrt{\rho_B}}\right] \label{Ec:FB}
\end{equation}
and
\begin{equation}
\mathbf{F}_F = - \nabla\left[A\,\rho_F^{2/3}
-\frac{\hbar^2}{6\,m_F}\, \frac{\nabla^2\sqrt{\rho_F}}{\sqrt{\rho_F}}
\right].
\label{Ec:FF}
\end{equation}
In these equations we shall set $\rho_{{\sigma}}(\mathbf
r,t)=\rho_{{\sigma}}(r)+\delta \rho_{{\sigma}}(\mathbf r,t)$ and
proceed first to discuss the equilibrium profiles $\rho_{{\sigma}}(r)$.

\section{Equilibrium  profiles and spatial separation}
\label{sec:equil}

The particle density profiles at equilibrium are obtained by imposing
the steady-state condition $\partial\mathbf{j}_{\sigma}/\partial t =0$
in Eq.~(\ref{Ec:finalsolution}). This ensures consistence between
static and dynamical solutions as well as fulfilment of the
generalized Kohn theorem \cite{kohn}. The above
equilibrium condition is in fact equivalent to a minimization of the
mean-field energy functional
\begin{eqnarray}
E[\rho_F,\rho_B]&=&\int\,d^3r\,\left(V_B\,\rho_B +
\frac{g_{BB}}{2}\,\rho_B^2 + \xi_B\right)\nonumber \\
&+&\int\,d^3r\,\left( V_F\,\rho_F +\frac{3}{5}\,A\,\rho_F^{5/3}+
\xi_F\right)\nonumber \\ &+& g_{BF} \int\,d^3r\,\rho_F\rho_B,
\label{Ec:functional}
\end{eqnarray}
where the quantum pressure or surface energy
terms read 
\begin{equation} 
\xi_B =
\frac{\hbar^2}{2\, m_B}|\nabla\sqrt{\rho_B}|^2 
\label{Ec:surfB}
\end{equation}
and
\begin{equation}
\xi_F = \frac{\hbar^2}{6\, m_F}|\nabla\sqrt{\rho_F}|^2.
\label{Ec:surfF}
\end{equation} 

As was found in previous studies \cite{Molmer_mix,
Viverit2000a, Minguzzi00a,Miyakawa00b}, on increasing the
boson-fermion repulsion the mixture undergoes spatial separation. In a
finite cloud the transition is smooth and can be described by
following the behavior of the boson-fermion interaction energy
$E_{int}=g_{BF} \int d^3 r\, \rho_B \rho_F$, which for a given
coupling strength is determined by the overlap of the two species
\cite{Akdeniz02a,Akdeniz02b}. The onset of the transition is signalled
first by a decrease of $E_{int}$ as a function of the boson-fermion
coupling (partial separation), until the value $E_{int}$ becomes
negligible (full demixing). The maximum of the interaction energy as a
function of $g_{BF}$ has been estimated for harmonic confinement at
zero temperature within the Thomas-Fermi approximation to lie at
\begin{equation}
\frac{a_{BF}^{part}}{a_{BB}} = \left(c_1\,\frac{N_F^{1/2}}{N_B^{2/5}} +
c_2\,\frac{N_B^{2/5}}{N_F^{1/3}}\right)^{-1},
\label{Ec:partdemix}
\end{equation}
where 
\begin{eqnarray}
c_1 &=& \frac{15^{3/5}}{48^{1/2}}\,\frac{(m_F+m_B)\,m_F^{1/2}}{2\,\,
m_B^{3/2}}\, \left(\frac{a_{BB}}{d}\right)^{3/5} \nonumber \\ c_2 &=&
\frac{48^{1/3}}{15^{3/5}}\,\left(\frac{6}{\pi}\right)^{2/3} \,
\frac{m_F+m_B}{2\,m_F}\left(\frac{a_{BB}}{d}\right)^{2/5}
\end{eqnarray}
and $d=(\hbar/m_B\omega_B)^{1/2}$ is the  bosonic harmonic-oscillator length.
The point of full demixing  as given by the Thomas-Fermi approximation
occurs instead at 
\begin{equation}
a_{BF}^{full} = \left(\dfrac{a_{BB}}{\alpha\,k_F}\right)^{1/2}
\label{Ec:fulldemix}
\end{equation}
where $k_F=(48N_F)^{1/6}/d$ and
$\alpha=[3^{1/3}/(2\pi)^{2/3}](m_F+m_B)^{2}/(4m_F m_B)$.  As we shall
see below, the analysis of the mode frequencies as functions of the
boson-fermion coupling strength yields a dynamical condition for
spatial  separation which is intermediate between the two static
conditions laid out in Eqs.\ (\ref{Ec:partdemix}) and (\ref{Ec:fulldemix}).

\section{Collective excitation spectrum}
\label{sec:colle}
We proceed to evaluate  the dynamical behavior of the cloud as it undergoes
spatial separation. Under the assumption of a weak external
drive  we can neglect anharmonic contributions. The
spectrum of collective excitations is then obtained by linearizing
 Eq.~(\ref{Ec:finalsolution}) around the equilibrium state. By taking the
Fourier transform with respect to the time variable we thus obtain
coupled eigenvalue equations for the density fluctuations
$\delta\rho_{\sigma}$ of each species,
\begin{equation} 
m_{\sigma}\,\Omega^2\,\delta\rho_{\sigma} =
\nabla\cdot\left(\rho_{\sigma}\, \delta\mathbf{F}_{\sigma}\right)
-g_{BF}\,\nabla\cdot\left(\rho_{\sigma}\,\nabla\delta\rho_{\bar{\sigma}}\right)\, .
\label{Ec:tosolve}
\end{equation} The expressions for the linearized forces 
$\delta\mathbf{F}_{\sigma}$ are given in Appendix \ref{app:forces}.

We have solved Eq.~(\ref{Ec:tosolve}) in the case of a spherically
symmetric confinement, setting the values of the trapping frequencies
at $\omega_F=\omega_B = \omega_0$ with $\omega_0 = 2\pi\times
1000\,$s$^{-1}$ and the values of the boson-boson scattering length at
$a_{BB} = 0.27\,$nm, which correspond to the $^6$Li-$^7$Li mixture in
the Paris experiment \cite{Schreck2001a,notina}.  We vary the mutual
scattering length $a_{BF}$ and the number of particles of each species
in order to explore the transition from the mixed state to the fully
separated state. In particular, the choice $N_F=10^4$ and
$N_B=2.4\times10^7$ leads to bosonic and fermionic clouds having
approximately the same size at zero coupling.

The numerical procedure that we have used can be summarized as
follows: ({\it i}) we look for a spherically symmetric 
steady-state solution of
Eqs.\ (\ref{Ec:finalsolution})-(\ref{Ec:FF}) by a
steepest-descent method 
\cite{Press92,Dalfovo1996a};
({\it ii}) we decompose the density fluctuations into components of
definite angular momentum, {\it i.e.} we factorize the amplitude of
the fluctuations as $\drho_{\sigma}(\mathbf{r}) =
\drho_{\sigma}^{l}(r)\,Y_{lm}(\hat{r})$ with $Y_{lm}$ the spherical
harmonic functions; and ({\it iii}) we set up an eigenvalue problem
for each $l$ by discretizing Eqs.\ (\ref{Ec:tosolve}) and  solve
it by means of  standard routines from the \texttt{LAPACK} library
\cite{lapack3}.

\begin{figure*}
\begin{tabular}{c}
\includegraphics[width=0.5\linewidth]{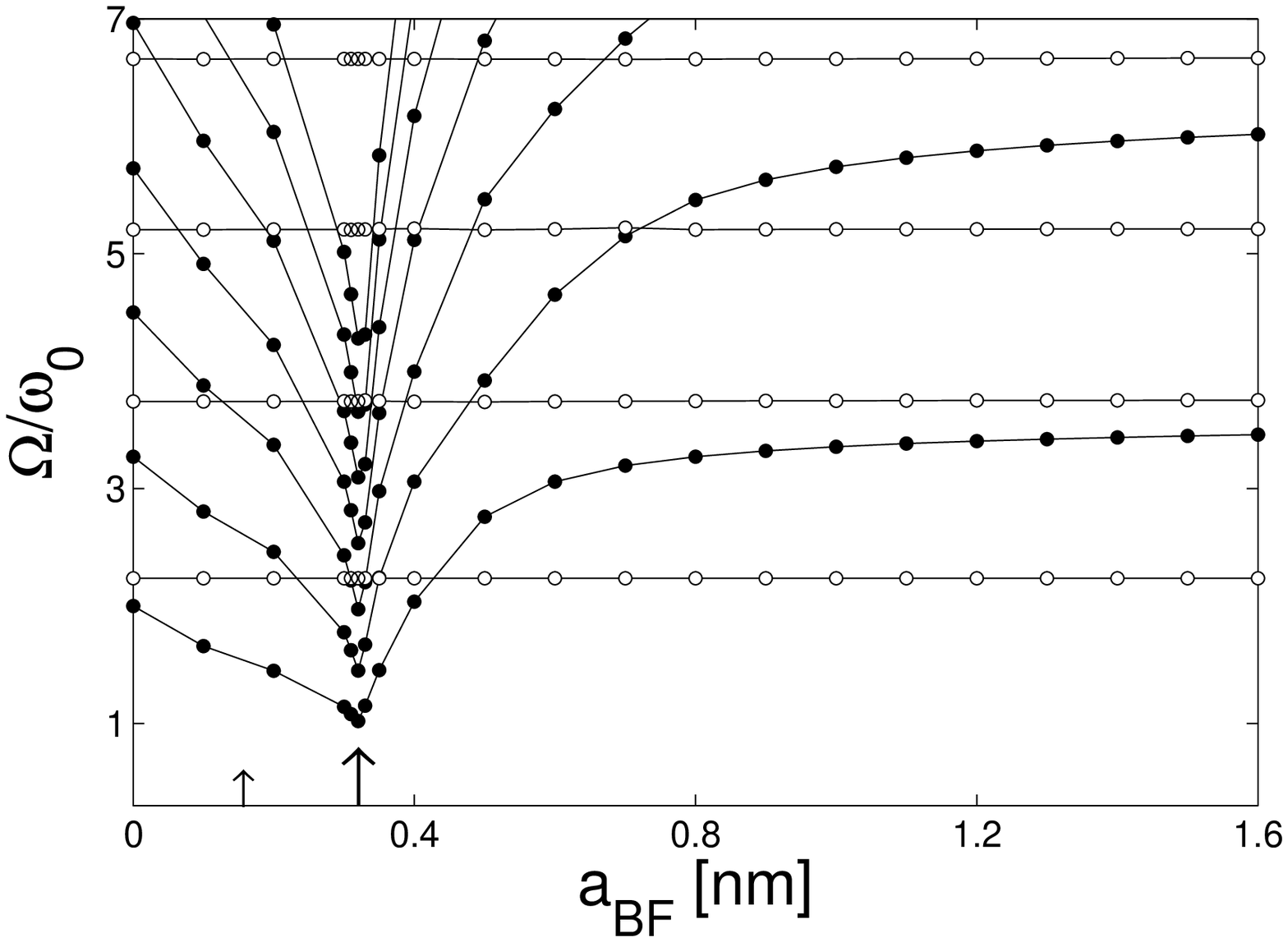} \\
\includegraphics[width=0.5\linewidth]{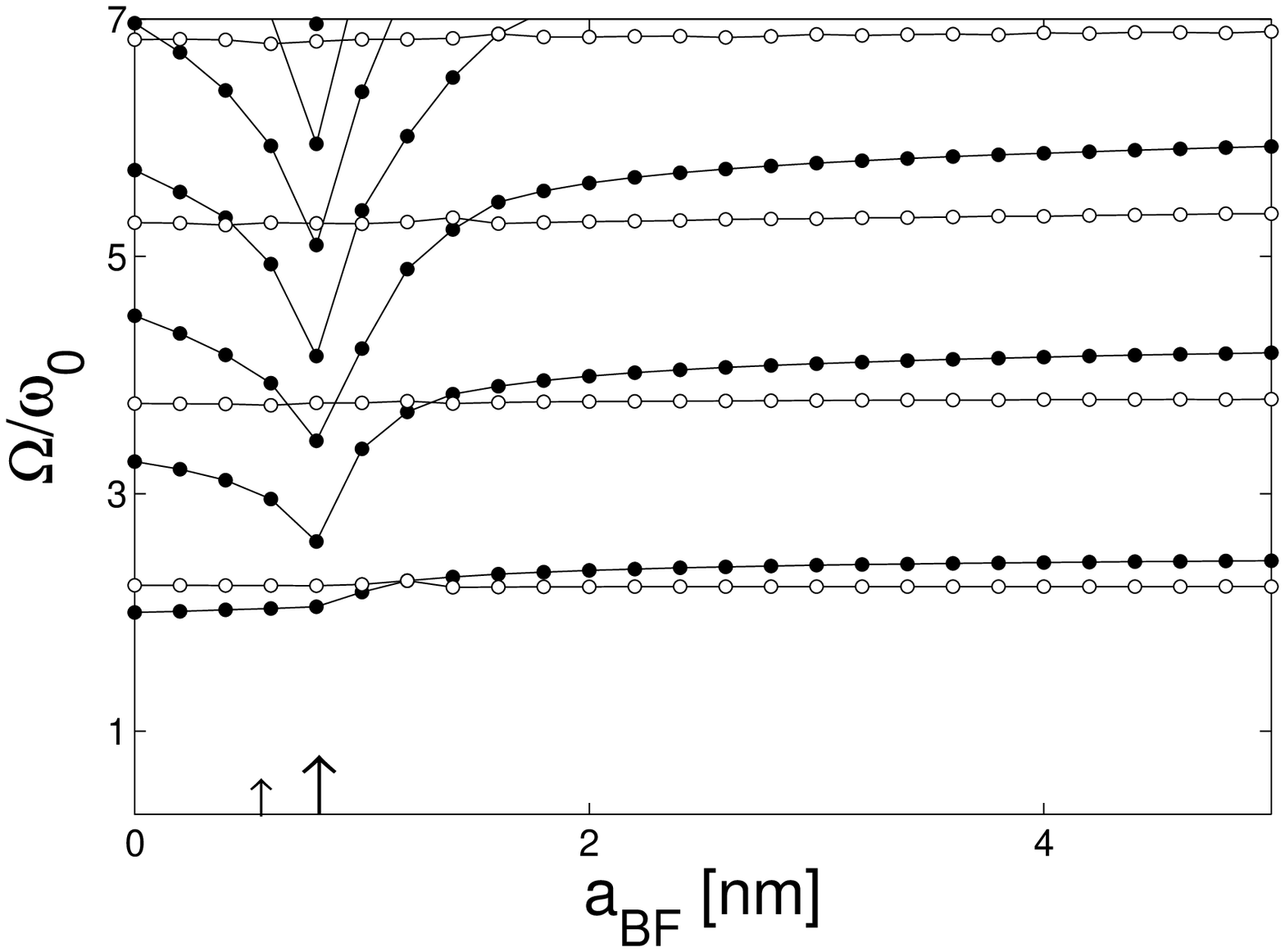} \\
\includegraphics[width=0.5\linewidth]{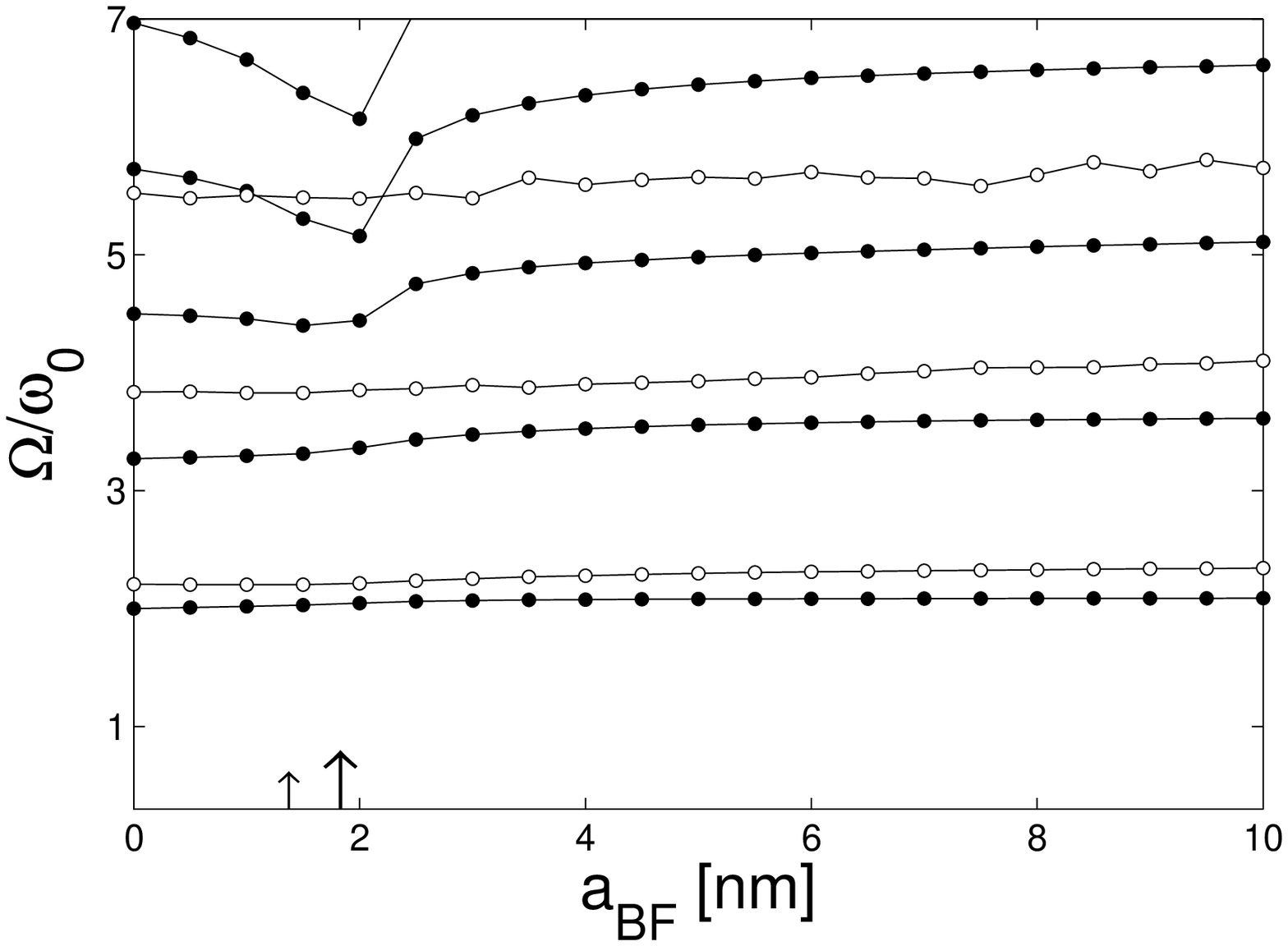} \\
\end{tabular}
\caption{\label{fig:spec0} Frequencies of $l=0$ collective modes (in
units of the trap frequency $\omega_0$) as functions of the
boson-fermion scattering length $a_{BF}$ (in nm). The panels from top
to bottom correspond to $N_B=2.4\times 10^7, 10^6$, and $ 10^5$
bosons.  Each frame displays the low-lying hydrodynamic modes
(fermionic -- dots and bosonic -- open circles) from the numerical
solution of Eq.~(\ref{Ec:tosolve}) with $N_F=10^4$ fermions and the
other parameters as specified in the text.  In each panel the small
arrow indicates the onset of partial demixing from a static criterion
(Eq.~(\ref{Ec:partdemix})) and the large arrow indicates the point of
vanishing fermionic density at the center of the trap
(Eq.~(\ref{Ec:sounddemix})). The lines are a guide to the eye.}
\end{figure*}

\begin{figure*}
\includegraphics[width=0.5\linewidth,clip=]{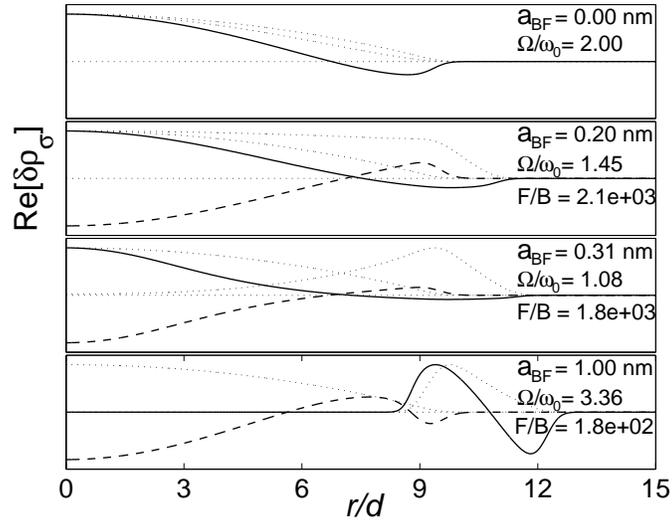} 
\caption{\label{fig:drho0} Monopolar density fluctuation profiles
  (arbitrary units) as
  functions of the radial coordinate (in units of the harmonic
  oscillator length $d$) for various values of  the boson-fermion
  coupling as indicated in the panels, with
  $N_B=2.4\times 10^7$ and $N_F=10^4$. Continuous and dashed  lines
  display the fermionic and bosonic profiles, respectively. The
  scale of each fluctuation has been changed to fit in the
  graph ($F/B$ indicates the ratio between the maximum value attained by
  the fermionic and bosonic fluctuation profiles).
  The dotted lines show the equilibrium density profiles, again rescaled
  to fit in the graph.}
\end{figure*}

The results for the frequencies of the monopole ($l=0$) modes are
given in Fig.\ \ref{fig:spec0} as functions of the boson-fermion
coupling for $N_F=10^4$ and various values of $N_B$.  A sampling of
the eigenvectors is reported in Fig.\ \ref{fig:drho0}. We have labelled the
modes as ``fermionic'' (dots) or ``bosonic'' (circles) according to
the nature of their eigenvalue 
 in the limit of vanishing $g_{BF}$. Of course, at
finite values of $g_{BF}$ the modes are coupled and the labels are
just conventional, but can still be assigned by looking at the nodes
of each density fluctuation: fermionic (bosonic) modes keep 
a constant number of nodes in $\delta\rho_F$ ($\delta\rho_B$) with
increasing $g_{BF}$. 

A common feature of our results in Fig.\ \ref{fig:spec0} is a
non-monotonic behavior of the frequency of the fermionic modes as the
cloud evolves from the mixed regime to the fully separated one. With
increasing $g_{BF}$ a fermionic mode acquires the character of an
out-of-phase fluctuation of the two components (see
Fig.~\ref{fig:drho0}) and we observe a softening of its frequency.
This we interpret as a signal of the approaching spatial-separation
transition: with increasing boson-fermion repulsion the out-of-phase
oscillation requires less and less energy, until at the transition the
mixture takes as its equilibrium configuration a state which
corresponds to a ``frozen'' out-of-phase oscillation. The nonzero
value of the lowest mode frequency at the transition point appears to
be due to the presence of the confinement, since in a linearized
theory it should tend to zero in the proper thermodynamic limit.

The transition point as dynamically determined by the sharp upturn in
the fermionic mode frequencies does not agree with the static
criterion (\ref{Ec:partdemix}) or (\ref{Ec:fulldemix}), but
corresponds to an intermediate point where the equilibrium density of
the fermionic cloud vanishes at the center of the trap (see the dotted
lines in Fig.~\ref{fig:drho0}). In the limit of large number of bosons
this point can be analytically estimated within the Thomas-Fermi
approximation to be
\begin{equation}
\frac{a_{BF}^{dyn}}{a_{BB}}=\frac{2m_F}{m_B+m_F} \frac{\mu_F}{\mu_B}
\label{Ec:sounddemix}
\end{equation}
where $\mu_F$ and $\mu_B$ are the chemical potentials for
fermions and bosons, respectively.  In a local-density picture this
corresponds to a minimum in the velocity of the fermionic sound wave.
The locations of the transition point according to the criterion 
in Eqs.~(\ref{Ec:partdemix}) and (\ref{Ec:sounddemix}) are shown by
the arrows in Fig.~\ref{fig:spec0}.

\begin{figure*}
\includegraphics[width=0.5\linewidth,clip=true]{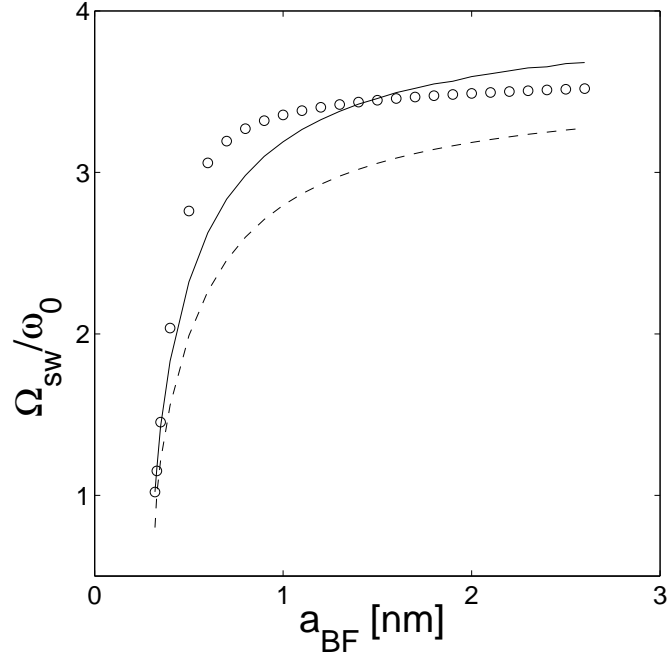}
\caption{\label{fig:sound2}
Frequency of the lowest fermionic monopole  mode after demixing,
as a function of $a_{BF}$ (in nm)  for $N_B=2.4\times
10^7$ and $N_F=10^4$. The circles are from the numerical solution of
Eq.~(\ref{Ec:tosolve}), while the lines show the predictions of Eq.\
(\ref{Ec:sw}) taking as effective density the maximum value of the
 equilibrium density (full line) or the density of a
uniform shell with $N_F=10^4$ fermions  inside the
Thomas-Fermi radii (dashed line).}
\end{figure*}

On further increasing the  boson-fermion coupling in the
spatially separated regime we observe that the frequency of the
fermionic modes continues to increase. This can be understood
 by means of a simple model
of sound-wave propagation inside a uniform shell of given thickness
$\eta$. The velocity $c_s$ of the wave is related to the
compressibility $\kappa$ of the fermionic gas according to
$c_s= (m_F\rho_F\,\kappa)^{-1/2}$ and  by imposing rigid boundary conditions at
the edges of the shell we can  obtain its frequency spectrum.  For
$l=0$ the lowest sound mode is given by \cite{note2}
\begin{equation}
 \Omega_{\text{SW}} = \frac{\pi c_s}{\eta}
\label{Ec:sw}
\end{equation} 
and  the sound velocity can be estimated from the ideal-gas expression of
the pressure ($p=\frac{2}{5}\,A\,\rho_F^{5/3}$) as
\begin{equation}
c_s^2=\frac{2}{3}\frac{A}{m_F}\,\rho_F^{2/3}. 
\end{equation}
Since we are assuming a constant-density shell for each value of
$a_{BF}$, to actually compare the predictions of Eq.~(\ref{Ec:sw})
with the numerical results in Fig.~\ref{fig:spec0} we need a suitable
choice of the effective fermionic density $\rho_F$.  In Fig.\
\ref{fig:sound2} we show the frequency of the lowest sound mode
calculated from Eq.~(\ref{Ec:sw}) as a function of $a_{BF}$ using two
choices for the effective density, one corresponding to the maximum
value of the fermionic equilibrium profile and the other evaluated for
a uniform spherical shell with the same number of fermions and the
radii taken from the Thomas-Fermi equilibrium profile.  In view of the
rough approximations that we are making, the agreement in Fig.\
\ref{fig:sound2} between the numerical calculation of the lowest
monopole frequency and this simple model is quite satisfactory.  Such
a degree of quantitative agreement is not found for the high-frequency
modes, since the density fluctuation tails become more marked and are
more sensitive to the inhomogeneity of the fermionic shell.

Returning to  Fig.\ \ref{fig:spec0}, it also shows that  the frequencies of the
bosonic monopole modes  are essentially unaffected by the boson-fermion
interaction. This is due to the fact that with our choice of system
parameters the bosonic cloud has a considerably higher density than
the fermionic one, leading to a weakly coupled dynamics. 
It is worth recalling that this is  at present a relevant
experimental situation, with the fermionic cloud
containing  a denser Bose-Einstein condensate.

\begin{figure*}
\begin{tabular}{cc}
$l=0$ & $l=1$ \\
\includegraphics[height=6.5cm,clip=]{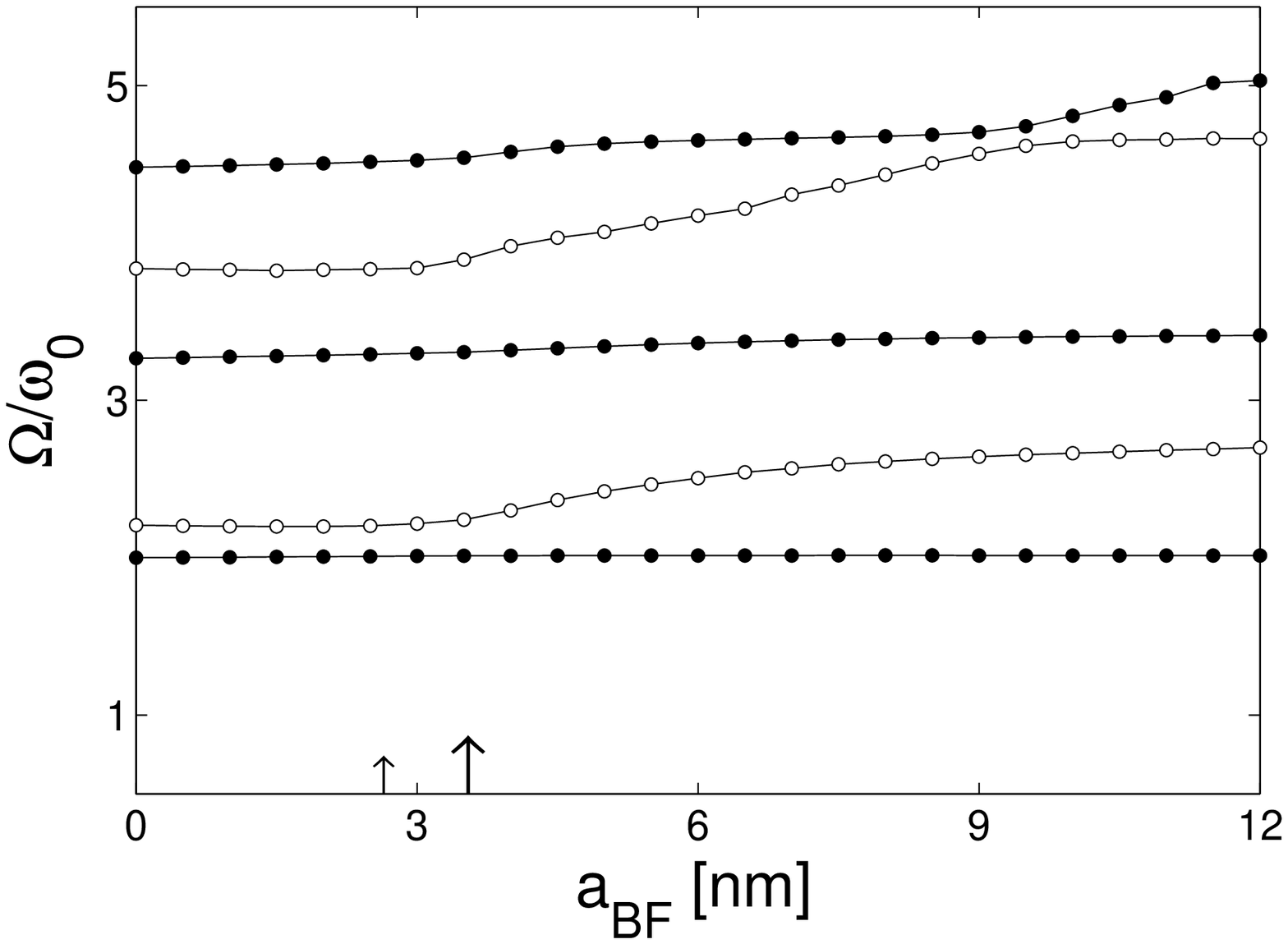} &
\includegraphics[height=6.5cm,clip=]{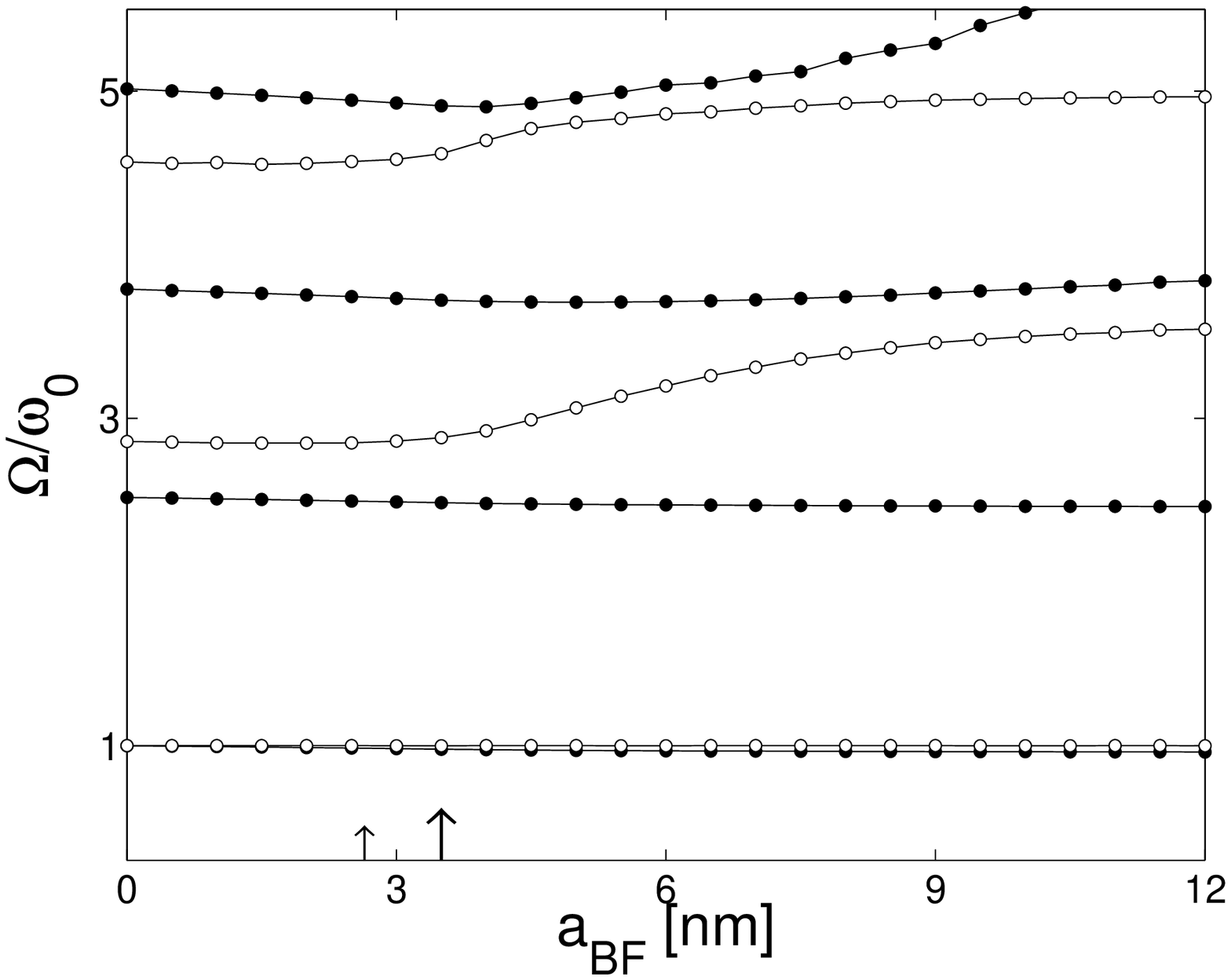}
\end{tabular}
\caption{\label{fig:highspec} Frequencies of the  $l=0$ (left panel) and
$l=1$ (right panel) collective excitations (in units of the trap
frequency $\omega_0$) as functions of $a_{BF}$ (in nm)
for the case $N_B=N_F=10^5$. The symbols and the system parameters are
 as in 
Fig.~\ref{fig:spec0}.}
\end{figure*}

We have also examined the case of comparable numbers of bosons and
fermions (see Fig.~\ref{fig:highspec}). In this case the frequency of
the low-lying bosonic monopole modes increases slowly with increasing
boson-fermion coupling while the frequency of the fermionic ones is
quite unchanged except for level crossings (left panel in Fig.\
\ref{fig:highspec}). The dipolar ($l=1$) modes show similar features
(right panel).  The behavior found for the higher modes, {\it e.g.}
those with frequencies larger than approximately $7\,\omega_0$, is
instead similar to that shown in Fig.\ \ref{fig:spec0} and is
therefore not reported in Fig.\ \ref{fig:highspec}.  These features
can be understood by considering the relative size of the two clouds:
the hole produced in the fermionic cloud by the bosons is in this case
so small that the low-lying fermionic fluctuations cannot sense it.

\begin{figure*}
\includegraphics[width=0.55\linewidth,clip=]{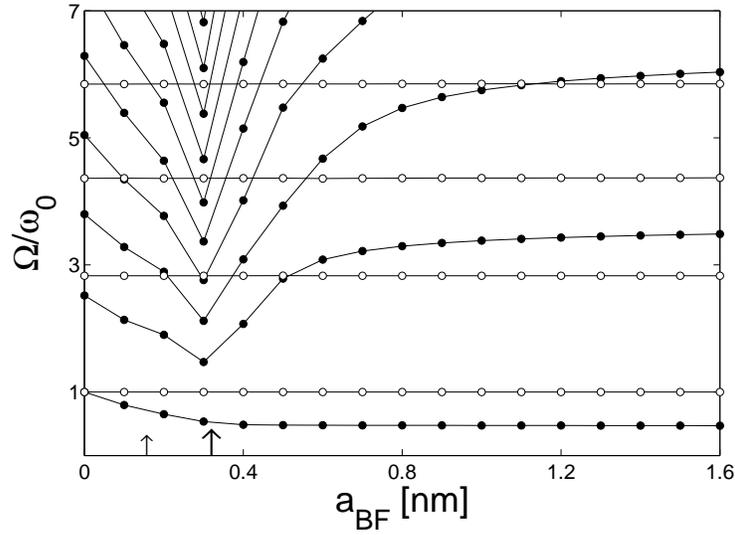}
\caption{\label{fig:spec1} Frequencies of the $l=1$ modes (in units of
 the trap frequency $\omega_0$) as functions of $a_{BF}$ (in nm) for
 $N_B=2.4\times 10^7$ and $N_F=10^4$.  The symbols 
 and the system parameters are as in Fig.\ \ref{fig:spec0}.}
\end{figure*}

Finally, we have solved the hydrodynamic equations for the $l=1$
dipolar  oscillations with a choice of particle numbers analogous to
that made for the monopole modes in Fig.~\ref{fig:spec0}. As  is shown in
Fig.~\ref{fig:spec1}, in this case the lowest fermionic-mode frequency
initially decreases 
monotonically and tends to a constant value for large values of
$g_{BF}$. In addition, we find the  Kohn mode 
at the frequency of the  trap for any value of the
boson-fermion coupling. The higher modes exhibit instead the
same behavior as the monopole modes in Fig.~\ref{fig:spec0}.

\section{Summary and concluding remarks}
\label{sec:summary} 

In summary, in this paper we have studied how the mutual repulsive
interactions affect the spectrum of collective excitations in a
trapped boson-fermion mixture in the collisional regime as the mixture
undergoes spatial separation. For this purpose we have derived and
solved the equations of generalized hydrodynamics beyond the
Thomas-Fermi approximation, by including surface density-gradient 
terms in the form first proposed by  von
Weizs\"acker.  When the two component clouds have similar sizes 
 (implying $N_B\gg N_F$), we have
found that the frequencies of the fermionic $l=0$ modes decrease as the
boson-fermion scattering length is increased and the mixture approaches
demixing. This frequency softening is directly related to the change
in shape of the equilibrium density profiles as the repulsive
interactions become stronger and reflects the tendency of the
two components  to spatially separate. At the point where the fermions are
expelled from the center of the trap, although the two clouds still
partially overlap, the fermionic mode frequencies  start
to grow in a fashion which essentially agrees  with a simple model of
sound wave propagation inside
a fermionic spherical shell.   A similar trend is also found in the
frequencies of the fermionic
$l=1$ bulk modes, while the bosonic dipolar surface mode displays the
Kohn-theorem behavior and is therefore unaffected by the interactions.
All these features are very different from those that we observe for
$N_B \simeq N_F$, when the bosonic modes are the most sensitive to the
boson-fermion coupling.

We have thus found a clear dynamical signature of the onset of spatial
demixing in the spectrum of collective modes in the case $N_B \gg
N_F$. This is expected to be helpful since the formation of a
symmetric ``egg'' configuration in the demixed cloud is not easily
detected from an analysis of column density profiles
\cite{Akdeniz02a}. The dynamical condition for demixing is given by
the point where the topology of the fermionic equilibrium density
profile changes as the fermions start to arrange themselves in a shell
around the bosons.  This point does not coincide with the points of
partial or full demixing as obtained from the static study of the
boson-fermion energy functional. However, this is easily understood if
one considers that in a finite system the transition is smooth and can
be characterized by several different conditions, which will coincide
only in the thermodynamic limit.

The present analysis can be extended to study the transition towards
``exotic'' configurations of the demixed cloud, as predicted by the
static study of the energy functional \cite{Akdeniz02b}.  A
calculation of the spectrum of collective excitations of a
boson-fermion mixture across spatial separation in the collisionless
regime is in progress and will be reported elsewhere.

\acknowledgments
This work was supported by INFM through the PRA-Photonmatter Program.

\appendix

\section{Full expressions for  force fluctuations}
\label{app:forces}
As was mentioned in the text, we decompose the particle density fluctuations
in their angular-momentum components as
$\delta\rho_\sigma(\mathbf{r})=\delta\rho_\sigma (r)\,Y_{lm}(\hat{r})$. Then, for
each $l$ we calculate $\delta\mathbf{F}_{\sigma} =
\mathbf{F}_{\sigma}[\rho_{\sigma}(r) +
\delta\rho_{\sigma}(r)\,Y_{lm}(\hat{r})]
-\mathbf{F}_{\sigma}[\rho_{\sigma}(r)] $ to linear terms in the
density fluctuations. This yields
\begin{widetext}
\bea \delta\mathbf{F}_{F}&=&
-\nabla\left\{\left[\frac{2}{3}\,A\,\rho_F^{-1/3}\,\delta\rho_F +
\frac{\hbar^2}{6\,m_F}\left(\frac{\rho_F'}{r\,\rho_F^2} +
\frac{\rho_F''}{2\rho_F^2} -
\frac{\rho_F'^2}{2\rho_F^3}+\frac{l(l+1)}{r^2\,\rho_F}
\right)\,\delta\rho_F \right.\right.  \nonumber \\
&&\left.\left. -\frac{\hbar^2}{6\,m_F}\left(\frac{1}{r\,\rho_F} -
\frac{\rho_F'}{2\,\rho_F^2} \right)\delta\rho_F'
-\frac{\hbar^2}{12\,m_F\rho_F} \,
\delta\rho_F''\right]Y_{l\,m} \right\}
\eea
and 
\bea
\delta\mathbf{F}_{B}&=&-\nabla\left\{\left[ g_{BB}\,\delta\rho_B +
\frac{\hbar^2}{2\,m_B}\,\left(\frac{\rho_B'}{r\,\rho_B^2} +
\frac{\rho_B''}{2\rho_B^2} - 
\frac{\rho_B'^2}{2\rho_B^3}+\frac{l(l+1)}{r^2\,\rho_B}\right)\delta\rho_B
\right.\right. \nonumber \\
&&\left. \left. -\frac{\hbar^2}{2\,m_B}\left(\frac{1}{r\,\rho_B} -
\frac{\rho_B'}{2\,\rho_B^2} \right)\delta\rho_B' -
\frac{\hbar^2}{4\,m_B} \frac{1}{\rho_B}\,\delta\rho_B'' \right]
Y_{lm}\right\}.
\eea 
\end{widetext}
In these equations a prime means a derivative with respect to
$r$.

\begin{thebibliography}{32}
\bibitem{notina}
Regarding the choice of the traps we have set the
bosonic and fermionic frequencies equal, taking the
geometrical average of the experimental values in the three
Cartesian directions.
\bibitem{note2}{This is readily
obtained by imposing rigid boundary conditions on the lowest $l=0$
solution for sound-wave propagation inside a uniform spherical shell
with hard walls. The wave is described by a   spherical Bessel
function $j_0$ of appropriate argument.}
\end{thebibliography}
\end{document}